\documentclass[letterpaper,11pt]{article}
\pdfoutput=1 

\usepackage{jcappub} 

\usepackage[T1]{fontenc} 

\def\beq{\begin{equation}}
\def\eeq{\end{equation}}

\title{Reheating constraints and consistency relations of the Starobinsky model and some of its generalizations}

\author[a]{Marcos A. G. Garcia,}
\author[b]{Gabriel Germ\'an,}
\author[b]{R.~Gonzalez~Quaglia}
\author[b,c]{and A.~M.~Moran~Colorado}

\affiliation[a]{Departamento de F\'isica Te\'orica, Instituto de F\'isica,\\
Universidad Nacional Aut\'onoma de M\'exico, Ciudad de M\'exico C.P. 04510, Mexico}
\affiliation[b]{Instituto de Ciencias F\'{\i}sicas, Universidad Nacional
Aut\'onoma de M\'exico,\\ Av. Universidad s/n, Cuernavaca, Morelos, 62210, Mexico}
\affiliation[c]{Centro de Investigaci\'on en Ciencias, Universidad Aut\'onoma del Estado de Morelos,\\ Avenida Universidad 1001, 
Cuernavaca, Morelos 62209, Mexico}

\emailAdd{marcos.garcia@fisica.unam.mx}
\emailAdd{gabriel@icf.unam.mx}
\emailAdd{rodrigo@icf.unam.mx}
\emailAdd{abril.moranclo@uaem.edu.mx}

\abstract{Building on the success of the Starobinsky model in describing the inflationary period of the universe, we investigate two simple generalizations of this model and their constraints imposed by the reheating epoch. The first generalization takes the form $R^{2p}$, while the second is the $\alpha$-Starobinsky model. We first focus on the case where $p=1$ or equivalently, $\alpha=1$, which corresponds to the original Starobinsky model.
We derive exact consistency relations between observables and cosmological quantities, without neglecting any terms, and impose the reheating condition $0 < \omega_{re} < 0.25$, where $\omega_{re}$ is the equation of state parameter at the end of reheating. This allows us to obtain new bounds for $n_s$ and $r$ that satisfy this condition and apply them to other observables and cosmological quantities. We repeat this process for the cases where $p \neq 1$ and $\alpha \neq 1$ and find that these generalizations only result in minor modifications of the Starobinsky model, including the potential and the bounds on observables and cosmological quantities. }

\begin{document}
\maketitle
\flushbottom

\section{Introduction}\label{Intro}

Cosmic inflation is characterized by an early-time accelerated expansion of the universe, typically modeled using a scalar field called the inflaton. However, scenarios with multiple fields, higher-spin fields, or even geometric methods based on higher-curvature corrections of the Einstein-Hilbert action, have been shown to lead to successful inflation (for reviews on inflation see e.g., \cite{Linde:1984ir}-\cite{Martin:2013tda}). The Starobinsky model \cite{Starobinsky:1980te} is an example of a geometric modification of standard General Relativity, in which a quadratic term of the scalar curvature is added to the gravitational action. This model, first proposed before the cosmic inflation paradigm was properly established, is in remarkable agreement with the current measurements of the power spectrum of primordial curvature fluctuations, and the present bounds on the spectrum of primordial gravitational waves, as determined by Planck and BICEP/Keck collaborations~\cite{Akrami:2018odb,BICEP:2021xfz,Tristram:2021tvh}. Remarkably, the Starobinsky model, originally defined in the Jordan frame, is equivalent up to a conformal transformation to a single field model with an asymptotically flat potential. Moreover, upon defining the Standard Model fields as minimally coupled to gravity in the Jordan frame, the conformal transformation to the Einstein frame induces a coupling between these fields and the inflaton, providing a natural graceful-exit and reheating mechanism~\cite{Vilenkin:1985md,Faulkner:2006ub,Gorbunov:2010bn}. Further generalizations of the Starobinsky model have been proposed, including scenarios with non-quadratic higher curvature terms $R^{2p}$~\cite{Schmidt:1989zz,Maeda:1988ab,Muller:1989rp,Motohashi:2014tra,Renzi:2019ewp}, and supergravity-inspired deformations of the inflaton potential known as $\alpha$-Starobinsky models (or E-models)~\cite{Ellis:2013nxa,Kallosh:2013yoa,Ellis:2019bmm}.

In this paper, we investigate the connections between the inflationary observables in the Cosmic Microwave Background, and other relevant cosmological quantities, such as the post-inflationary reheating temperature, of the Starobinsky model and two of its generalizations. Our focus is on the Einstein frame, where the associated scalar field has a potential. By obtaining an explicit form for the value of the inflaton field at horizon crossing, we determine consistency relations between the tilt of the curvature power spectrum, $n_s$, the tensor-to-scalar ratio, $r$, and the corresponding runnings of the scalar and tensor indices, $n_{sk}$ and $n_{tk}$, respectively. Imposing the Planck bounds for $n_s$ \cite{Akrami:2018odb}, we establish constraints on all other relevant cosmological quantities, which we refer to as ``Planck constraints to inflation''.

Furthermore, we explore the implications of the Planck ranges for reheating scenarios (for reviews on reheating see e.g., \cite{Bassett:2005xm}-\cite{Amin:2014eta}) and investigate the effective equation of state parameter of reheating, $\omega_{re}$, constrained by the Planck data. This analysis sets weak bounds on possible reheating scenarios. We also investigate ``reheating constraints to inflation'' by constraining the value of $\omega_{re}$ and deriving new constraints for $n_s$, the other observables, and quantities such as the number of $e$-folds during inflation $N_{k}$, and reheating $N_{re}$. We also determine the relevant constraints for the $\alpha$-Starobinsky and $R^{2p}$ generalizations of the Starobinsky model.

The paper is organized as follows: In Section $\ref{Staro}$, we study the consistency relations and corresponding bounds for the Starobinsky model  \cite{Starobinsky:1980te}. In Section $\ref{GStaro}$, we introduce the $R^{2p}$ generalization of the model following~\cite{Motohashi:2014tra,Renzi:2019ewp}, and analyze the constraints, focusing on how the new parameter $p$ affects the results. We find that the generalization is less restrictive, but the allowed ranges of $p$ are very close to unity, corresponding to the original Starobinsky model. In Section $\ref{Ellis}$, we investigate the constraints for the $\alpha$-Starobinsky generalization \cite{Ellis:2013nxa,Kallosh:2013yoa,Ellis:2019bmm} and find that it only accounts for minor modifications to the original potential. Section $\ref{Con}$ concludes our paper, and a final appendix deals with the reheating temperature in the original model and its $p$-dependent generalization.

\section {The Starobinsky model}\label{Staro}

The Starobinsky model~\cite{Starobinsky:1980te} was one of the first proposed models for inflation, initially suggested prior to the establishment of the cosmic inflation paradigm. It is characterized by an action that includes both the Einstein-Hilbert term and a higher-order scalar curvature term
\begin{equation}
\label{Staropot}
S=\frac{M_{Pl}^{2}}{2}\int d^{4}x \sqrt{-g}\left(R+\frac{1}{6M^2} R^{2}\right),
\end{equation}
where $R$ is the Ricci scalar, $M_{Pl}=2.44\times 10^{18} \,\mathrm{GeV}$ is the reduced Planck mass, and $M$ is the Starobinsky free parameter, with mass units. Like any $f(R)$ theory, the Starobinsky action can be transformed into the form of the Einstein-Hilbert term plus an action for a scalar field. This form is achieved by performing the conformal transformation on the metric 
\begin{equation}
g_{\mu\nu}\rightarrow e^{\sqrt{\frac{2}{3}}\frac{\phi}{M_{Pl}}}g_{\mu\nu}.
\end{equation}
Applying this transformation to the Starobinsky model, we end up with the following action
\begin{equation}
S=\int d^{4}x \sqrt{-g}\left(\frac{M_{Pl}^{2}}{2}R-\frac{1}{2}\partial_{\mu}\phi\partial^{\mu}\phi-V(\phi)\right).
\end{equation}
Here $V(\phi)$ is the potential of the scalar field, given by
\begin{equation}
\label{Spot}
V(\phi)=V_{0}\left(1-e^{-\sqrt{\frac{2}{3}}\frac{\phi}{M_{Pl}}}\right)^{2},
\end{equation}
where $V_{0}\equiv \frac{3}{4}M_{Pl}^{2}M^2$.
With the model now cast in the Einstein frame and a scalar potential $V(\phi)$ identified, we can apply the standard expressions for slow-roll single-field inflation. This allows us to establish relationships between cosmological observables such as $r$, $n_{sk}$, $n_{tk}$, and the scalar spectral index $n_s$. Here, $n_{sk}$ is the running of the scalar spectral index, while $n_{tk}$ denotes the tensor running. After the consistency relations are found, we use Planck bounds on $n_{s}$ to set bounds on the other observables and cosmological quantities of interest.

\subsection{Consistency relations and Planck constraints to inflation}

Contact with models of inflation is achieved through cosmological observables which to first order in the slow-roll (SR) approximation are given by (see e.g.,  \cite{Lyth:1998xn}, \cite{Liddle:1994dx})
\begin{eqnarray}
n_{t} &=&-2\epsilon = -\frac{r}{8} , \label{Int} \\
n_{s} &=&1+2\eta -6\epsilon ,  \label{Ins} \\
n_{sk}\equiv \frac{d n_s}{d \ln k} &=&16\epsilon \eta -24\epsilon ^{2}-2\xi_2, \label{Insk} \\
n_{tk}\equiv \frac{d n_t}{d \ln k} &=&4\epsilon\left( \eta -2\epsilon\right), \label{Intk} \\
A_s(k) &=&\frac{1}{24\pi ^{2}} \frac{V}{\epsilon\, M_{Pl}^{4}}, \label{IA} 
\end{eqnarray}
where $n_t$ is the tensor spectral index, $r$ is the usual tensor-to-scalar ratio,  $n_s$ the scalar spectral index, $n_{sk}$ its running (usually denoted by $\alpha$) and $n_{tk}$ the running of the tensor index, in a self-explanatory notation. The amplitude of density perturbations at wave number $k$ is denoted by $A_s(k)$. All quantities are evaluated at horizon crossing at wavenumber $k$. The SR parameters appearing above are 
\beq
\epsilon \equiv \frac{M_{Pl}^{2}}{2}\left( \frac{V^{\prime }}{V }\right) ^{2},\quad\quad
\eta \equiv M_{Pl}^{2}\frac{V^{\prime \prime }}{V}, \quad\quad
\xi_2 \equiv M_{Pl}^{4}\frac{V^{\prime }V^{\prime \prime \prime }}{V^{2}},
\label{Spa}
\eeq
where primes on $V$ denote derivatives with respect to the inflaton $\phi$.

To find a relation between $n_{s}$ and the rest of the quantities of interest it is convenient to find a closed-form expression for the inflaton field at horizon crossing, $\phi=\phi_k$. This can be achieved by solving Eq.~(\ref{Int}), which yields
\beq
\phi_{k}=\sqrt{\frac{3}{2}}M_{Pl}\ln \left(1+\frac{8}{\sqrt{3r}}\right).
\eeq
With this, it is easy to find the relation between $r$ and $n_{s}$. In particular, Eq.~$(\ref{Ins})$ may be rewritten in the form 
\beq
\label{r1}
r=\frac{4}{3}\left(2-2\sqrt{1+3\delta_{n_s}}+3\delta_{n_s}\right),         \quad\quad \delta_{n_s}\equiv 1-n_s.
\eeq
Similarly, using the equations for $n_{sk}$ and $n_{tk}$  we find
\beq
\label{nsk1}
n_{sk}=\frac{1}{18}\left(4+9\delta_{n_s}-9\delta_{n_s}^2-\sqrt{1+3\delta_{n_s}}\left(4+3\delta_{n_s}\right)\right),
\eeq
\beq
\label{ntk1}
n_{tk}=\frac{1}{36}\left(8+3(4-3\delta_{n_s})\delta_{n_s}-8\sqrt{1+3\delta_{n_s}}\right).
\eeq
Eqs.~(\ref{r1}), (\ref{nsk1}) and (\ref{ntk1}) are the consistency relations for the Starobinsky model. Eqs.(\ref{nsk1}) and (\ref{ntk1}) can also be written in terms of $r$ as follows
\beq
\label{nsk2}
n_{sk}=-\frac{1}{96}r\left(16+10\sqrt{3r}+3r\right),
\eeq
\beq
\label{ntk2}
n_{tk}=-\frac{1}{192}r\left(8\sqrt{3r}+3r\right).
\eeq

\begin{figure}[t!]
\centering
  \includegraphics[width=.82\linewidth]{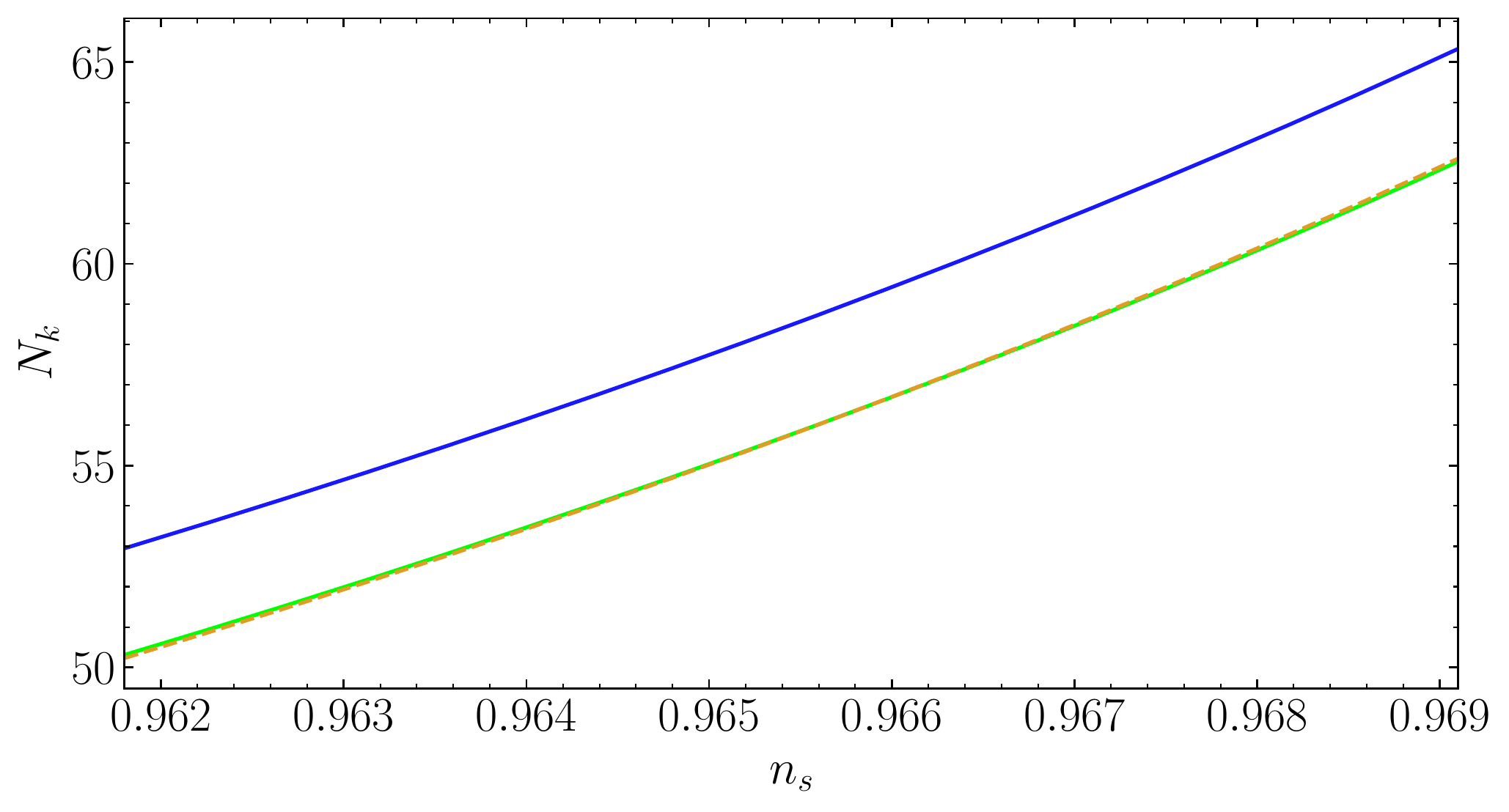}
  \caption{Plot of the number of $e$-folds during inflation $N_k$. The blue (upper) curve corresponds to neglecting the linear term in eq.~(\ref{Nk1}) while the green (lower) and dashed curves correspond to the exact formula and the approximate formula where $\phi_k$ is replaced by its mean value, respectively. Neglecting the linear term in Eq.~(\ref{Nk1}) corresponds to neglecting approximately 3 $e$-folds of inflation while approximating $\phi_k$ by its mean value perfectly overlaps with the exact result in the relevant range for $n_s$. }
  \label{fiaprox}
\end{figure}

Given the horizon exit value for the inflaton, we can determine the number of $e$-folds between horizon crossing and the end of inflation, $N_k$, by means of the SR approximation,
\beq
\label{Nk1}
 N_{k}=-\frac{1}{M_{Pl}^{2}}\int _{\phi_{k}}^{\phi_{e}} \frac{V}{V'} d\phi= \frac{3}{4}\left(e^{\sqrt{\frac{2}{3}}\frac{\phi_{k}}{M_{Pl}}}-e^{\sqrt{\frac{2}{3}}\frac{\phi_{e}}{M_{Pl}}}+\sqrt{\frac{2}{3}}\left(\frac{\phi_{e}-\phi_{k}}{M_{Pl}}\right)\right),
\eeq
where $\phi_{e}$ is the field evaluated at the end of inflation, which for the sake of analyticity we approximate as $\epsilon\simeq 1$,
\beq
\label{fie}
\phi_{e}=\sqrt{\frac{3}{2}}M_{Pl}\ln \left(1+\frac{2}{\sqrt{3}}\right).
\eeq
At the end of inflation, the SR approximation fails for the Starobinsky model but the small discrepancies from the value obtained in (\ref{fie}) give rise to negligible differences in the number of $e$-folds of inflation \footnote{In \cite{Ellis:2015pla}, a slightly smaller value of $\phi_e$ was considered, but this only leads to a difference of less than 0.2 $e$-folds in the resulting value of $N_k$ compared to the value presented here.}. Hence, $N_{k}$ may be approximated as
\beq
\label{Nk2}
N_{k}=\frac{\sqrt{3}}{2}\left(\frac{4}{\sqrt{r}}-1\right)+\frac{3}{4}\ln\left(\frac{\sqrt{3}+2}{\sqrt{3}+\frac{8}{\sqrt{r}}}\right).
\eeq
Typically, to obtain the consistency relations, one starts from an approximate version of $N_k$. Specifically, the linear term is neglected in (\ref{Nk1}): $\frac{3}{4}\sqrt{\frac{2}{3}}\left(\frac{\phi_{e}-\phi_{k}}{M_{pl}}\right)$, effectively making $\phi_{e}=\phi_{k}$ and then solving for $\phi_{k}$. This term is exactly $\frac{3}{4}\ln\left(\frac{\sqrt{3}+2}{\sqrt{3}+\frac{8}{\sqrt{r}}}\right)$ in (\ref{Nk2}) which, for the range of values of $r$ given in Table~\ref{boundsplin}, corresponds to neglecting almost 3 $e$-folds of inflation. Here we have not proceeded in this way obtaining with our approach exact consistency relations within the SR approximation. If we insist in obtaining $\phi_k$ using some approximate expression of $N_k$ it is much more convenient to approximate the linear term in $\phi_k$ by its mean value as shown in Fig.~\ref{fiaprox}. Thus, an expansion to first order of Eqs.~(\ref{r1})  and  (\ref{nsk2}) reduce to equations Eq.~(32) of \cite{Motohashi:2014tra}. Also, combining (\ref{Nk2}) with the consistency relations we get, to first order in $1/N_{k}$
\beq
\label{largeNk}
n_s=1-\frac{2}{N_k}, \quad\quad r=\frac{12}{N_k^2}, \quad\quad n_{sk}=-\frac{2}{N_k^2}, \quad\quad n_{tk}=-\frac{3}{N_k^3},
\eeq
the first three as in Eq.~(31) of \cite{Motohashi:2014tra}. Now that we have the consistency relations (\ref{r1}), (\ref{nsk1}), (\ref{ntk1}) and the formula for $N_k$ Eq.(\ref{Nk2}) we can find the bounds for each of these quantities by using the Planck range  on $n_{s}$ given in the fifth column of Table 3 of the Planck Collaboration paper~\cite{Akrami:2018odb}:\,\,  $0.9618<n_s<0.9698$.\footnote{We favor these constraints over those in~\cite{BICEP:2021xfz}, as the ranges derived in this reference assume a vanishing tensor spectral index $n_t=0$, incompatible with the consistency relation (\ref{Int}).}
\begin{table*}[t!]\label{Inflation bounds}
 \begin{center}
{\begin{tabular}{cccc}
\small
Quantity & Bounds & Consistency relation\\ \hline\\[0.1mm]
$n_s$   &     $0.9618 < n_{s} < 0.9698 $ & $----- $&\\[2mm] 
$r$   &   $0.00414> r >0.00262$ & $r=\frac{4}{3}\left(2-2\sqrt{1+3\delta_{n_s}}+3\delta_{n_s}\right)$\\[2mm]
$ n_{sk}$   &   $ -7.4\times 10^{-4} < n_{sk} < -4.6\times 10^{-4}$ & $n_{sk}=\frac{1}{18}\left(4+9\delta_{n_s}-9\delta_{n_s}^2-\sqrt{1+3\delta_{n_s}}\left(4+3\delta_{n_s}\right)\right)$\\[2mm]
$ n_{tk}$   &   $ -2.0\times 10^{-5} < n_{tk} < -9.8\times 10^{-6}$ & $n_{tk}=\frac{1}{36}\left(8+3(4-3\delta_{n_s})\delta_{n_s}-8\sqrt{1+3\delta_{n_s}}\right) $\\[2mm]
$N_{k}$   &   $50.3< N_{k} < 64.0$ &   $N_k=\frac{\sqrt{3}}{2}\left(\frac{4}{\sqrt{r}}-1\right)+\frac{3}{4}\ln\left(\frac{\sqrt{3}+2}{\sqrt{3}+\frac{8}{\sqrt{r}}}\right)$\\[2mm]
\end{tabular}}
\caption{\label{boundsplin} Bounds for the observables $r$, the scalar running $n_{sk}$, and the tensor running $n_{tk}$ obtained from the consistency relations for the Starobinsky model. The consistency relations are shown in the third column and given by Eqs.~(\ref{r1}), (\ref{nsk1}) and (\ref{ntk1}) respectively, where $\delta_{n_s}\equiv 1-n_{s}$. These bounds are calculated from the range for the spectral index $n_s$ as given in Table 3 of  \cite{Akrami:2018odb} . Although the number of $e$-folds during inflation $N_k$ is not an observable and therefore does not satisfy a consistency relation, bounds are also given here.}
\end{center}
\end{table*}

\subsection{Reheating constraints to inflation}

We now set constraints from reheating. To do this we need an expression for the EoS $\omega_{re}$ in terms of $n_s$. The number of $e$-folds of reheating $N_{re}\equiv \ln\left(\frac{a_r}{a_e}\right)$ can be written as
\beq
\label{Nre}
N_{re}=\frac{1}{3\left(1+\omega_{re}\right)}\ln \left(\frac{\rho_e}{\rho_{re}}\right).
\eeq
Here, $\omega_{re}$ is an effective equation of state assumed to be constant during reheating, $\rho_e = \frac{3}{2}V_e = \frac{9}{2}\frac{V_e}{V_k}H_k^2M_{Pl}^2$, and from the amplitude of scalar perturbations given by Eq.~(\ref{IA}), we have $H_k=\pi \sqrt{\frac{A_s r}{2}} M_{Pl}$. The energy density at the end of reheating is given by $\rho_{re}=\frac{\pi^2 g_{re}}{30}T_{re}^4$, where $g_{re}$ is the effective number of relativistic species. We can determine the reheating temperature $T_{re}$ from entropy conservation after reheating \cite{German:2022sjd}
\beq
\label{Nrp}
N_{rp}\equiv \ln\left(\frac{a_p}{a_r}\right)=\ln\left(\frac{a_p T_{re}}{\left(\frac{43}{11 g_{s,re}} \right)^{1/3}a_0T_0}\right),
\eeq
where $N_{rp}$ is the number of $e$-folds from the end of reheating  to the pivot scale with wavenumber $k_p$ and scale factor $a_p$, $g_{s,re}$ are the entropy degrees of freedom, $a_{0}$ and $T_{0}$ are the scale factor and temperature measured today.

\begin{figure}[t!]
\centering
  \includegraphics[width=.8\linewidth]{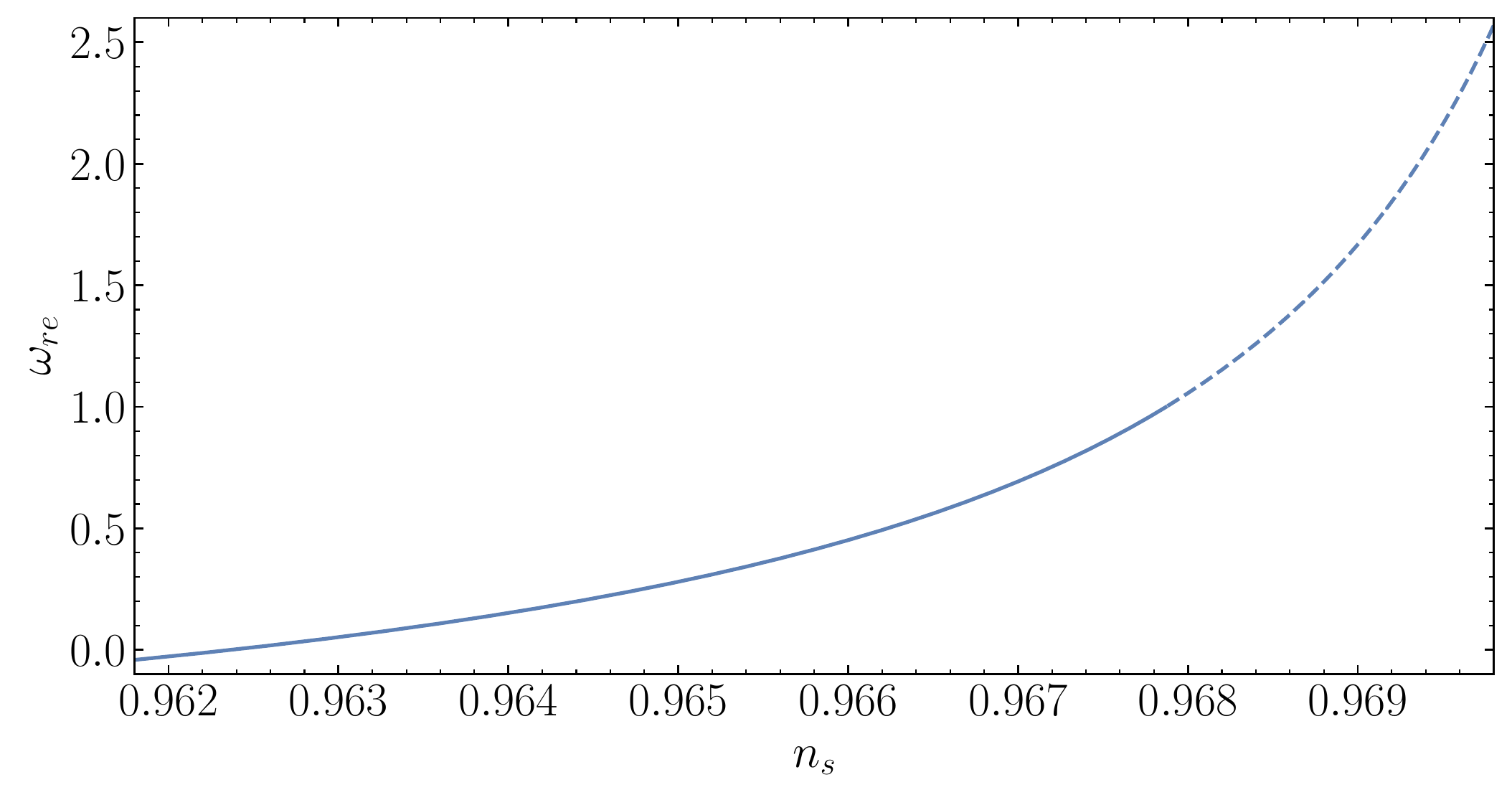}
  \caption{Plot of the equation of state parameter (EoS) during reheating as a function of $n_{s}$ where we considered the range $0.9618 < n_{s} < 0.9698$ \cite{Akrami:2018odb}. The section of the curve marked by dots represents a range of EoS values $\omega_{re}$ for which the authors are not aware of any existing models.}
  \label{omegare}
\end{figure}

As we cannot isolate $\omega_{re}$ using only the equations above and at the same time determine $N_{re}$ we consider the expansion from the time  a mode of wavelength $k$ exited the horizon during inflation to the time the mode reenters the horizon during radiation domination, at the pivot scale of wavenumber $k_p$ 
\beq
N_{kp}\equiv \ln\left(\frac{a_p}{a_k}\right)= \ln\left(\frac{a_e}{a_k}\right)+ \ln\left(\frac{a_r}{a_e}\right)+ \ln\left(\frac{a_p}{a_r}\right)=N_{k}+N_{re}+N_{rp}.
\eeq
Substituting $N_{re}$ above in Eq.~(\ref{Nre}) and solving for $\omega_{re}$ we get
\beq
\label{wre}
\omega_{re}=-1+\frac{1}{3} \frac{\ln\left(\rho_e/\rho_{re}\right)}{N_{kp}-N_k-N_{rp}}\,,
\eeq
where $N_{kp}\equiv \ln\left(\frac{a_p}{a_k}\right)=\ln\left(\frac{a_p H_k}{k_p}\right)=\ln \left(\frac{a_p\pi\sqrt{A_s r}}{\sqrt{2}k_p}\right)$ \cite{German:2020iwg}. Notice that now we have achieved the desired result, we have found an equation relating $\omega_{re}$ and $n_{s}$  however, we have a dependence on the reheating temperature $T_{re}$ inside of $N_{rp}$. In the case of a minimal coupling of the Standard Model to gravity in the Jordan frame, this temperature can be estimated to be $3.1\times 10^{9}\, {\rm GeV}$~\cite{Gorbunov:2010bn} (see Appendix). In order to evaluate the expression $(\ref{wre})$ we use the numerical values listed in Table~\ref{parameters}. 
\begin{table*}[t!]
\begin{center}
\begin{tabular}{ccc}
Parameter &\quad usually given as   &\,\, dimensionless, used here\\ \hline 
& \quad   & \quad  \\[-2mm]
$H_0$ & \quad $100\,h\ {\rm km/s/Mpc}$  & \quad $8.7581 \times 10^{-61}\,h$ \\[2mm]
$T_0$ & \quad $2.7255\, {\rm K}$  & \quad $9.6423\times 10^{-32}$\\[2mm]
$A_s$ & \quad $2.1 \times 10^{-9}$  & \quad $2.1 \times 10^{-9}$\\[2mm]
$k_p$ & \quad $0.05\ {\rm Mpc}^{-1}$  & \quad $1.3128\times 10^{-58}$\\[2mm]
$a_p$ & \quad $-$  & \quad $4.65\times 10^{-5}$ \\[2mm]
$a_{eq}$  & \quad $2.94\times 10^{-4}$  & \quad $1.33\times 10^{-4}h^{-2}$\\[2mm]
$a_{0}$  & \quad $-$  & \quad $1$\\[2mm]
$\Omega_{rd,0}$   & \quad $4.2\times 10^{-5}h^{-2}$  & \quad $4.2\times 10^{-5}h^{-2}$\\[2mm]
$\Omega_{md,0}$  & \quad $-$ & \quad $0.315$\\[2mm]
$h$  & \quad $-$  & \quad $0.674$\\[2mm]
$M_{pl}$  & \quad $2.44\times 10^{18}$ GeV  & \quad $1$\\[2mm]
 \end{tabular}
 \caption{\label{parameters} Numerical values used in the calculations. Dimensionless quantities are obtained by transforming them into units of $M_{pl}$ and then setting it equal to one. }
 \end{center}
\end{table*}
In Fig.~\ref{omegare} we show the behaviour of $\omega_{re}$ as a function of the spectral index $n_s$.
We now  impose constraints to inflation coming from reheating. For potentials of the form $V\propto \phi^{n}$, the equation of state parameter takes the form $\omega_{re}=\frac{n-2}{n+2}$ during the coherent oscillation of the field around its minimum ~\cite{Turner:1983he}. In the case of a quadratic potential, this leads to $\omega_{re}=0$. However, it is important to emphasize that this result is based on the assumption of perturbative decay of the inflaton into Standard Model particles.

After the inflationary phase, the equation of state parameter (EoS) averaged over the inflation field's oscillations is approximately $\omega_{re}=0$. Nevertheless, a significant transition occurs at the end of the preheating phase \cite{Dufaux:2006ee}. Within a time interval of about $10^{-36}$ seconds, the EoS undergoes a sharp change from $\omega_{re}=0$ to a value around $\omega_{re}\approx 0.2-0.3$, \cite{Podolsky:2005bw}. It is worth noting that the EoS does not immediately reach the radiation-dominated value of 1/3; instead, it settles around $\omega_{re}\approx 1/4$, as depicted in Fig. 1 of \cite{Podolsky:2005bw}, and remains at that value until complete thermalization is achieved.
This departure from an immediate radiation-dominated regime after preheating can be attributed to two factors. First, the light field retains a significant induced effective mass due to interactions. Second, there is a substantial residual contribution from the homogeneous inflaton field. These factors contribute to the deviation from an immediate transition to radiation dominance \cite{Podolsky:2005bw}. Therefore, it is reasonable to consider the range $0 < \omega_{re} < 0.25$ for the EoS during reheating \cite{Dai:2014jja}, \cite{Munoz:2014eqa}.  In table~\ref{boundsrein} bounds for the observables coming from the consistency relations eqs.~(\ref{r1}), (\ref{nsk1}) and (\ref{ntk1}) as well as for the number of $e$-folds during inflation $N_k$ and reheating $N_{re}$ are given. 
\begin{table*}[t!]
 \begin{center}
{\begin{tabular}{ccc}
\small
Quantity & Bounds & Defining equation\\ \hline\\[0.1mm]
$\omega_{re}$   &     $0 < \omega_{re} < 0.25 $ & Eq.~(\ref{wre})\\[2mm] 
$n_s$   &     $0.9623 < n_{s} < 0.9648 $ & Eq.~(\ref{Ins})\\[2mm] 
$r$   &   $0.0040> r > 0.0035$ &  Eq.~(\ref{r1})\\[2mm]
$ n_{sk}$   &   $ -7.2\times 10^{-4} < n_{sk} < -6.3\times 10^{-4}$ & Eq.~(\ref{nsk1})\\[2mm]
$ n_{tk}$   &   $ -1.9\times 10^{-5} < n_{tk} < -1.5\times 10^{-5}$ & Eq.~(\ref{ntk1})\\[2mm]
$N_{k}$   &   $51.0< N_{k} < 54.7$ & Eq.~(\ref{Nk2})\\[2mm]
$N_{re}$   &   $18.2> N_{re} > 14.6$ & Eq.~(\ref{Nre})\\[2mm]
\end{tabular}}
\caption{\label{boundsrein} Bounds for the observables $r$, the scalar running $n_{sk}$, and the tensor running $n_{tk}$ as well as for the number of $e$-folds during inflation $N_k$ and during reheating $N_{re}$. These bounds are obtained by imposing the range $0 < \omega_{re} < 0.25$ on the EoS during reheating and deriving from there new bounds for the spectral index $n_s$. The bounds for the observables are obtained  by using the consistency relations of the Starobinsky  model given by Eqs.~(\ref{r1}), (\ref{nsk1}) and (\ref{ntk1}).}
\end{center}
\end{table*}
\newpage

\section{\boldmath $R^{2p}$ Generalization of the Starobinsky model}\label{GStaro}

We now  study a generalization of the Starobinsky potential of the form
\beq
\label{genpot}
V_{g}= V_0e^{-2\sqrt{\frac{2}{3}}\frac{\phi}{M_{pl}}}\left(e^{\sqrt{\frac{2}{3}}\frac{\phi}{M_{pl}}}-1\right)^{\frac{2p}{2p-1}},
\eeq
where $V_{0}= \frac{2p-1}{2}\left(\frac{1}{2p}\right)^{\frac{2p}{2p-1}}\left(6M^2\right)^{\frac{1}{2p-1}}M_{Pl}^2$. This generalization of the Starobinsky model has been studied in various contexts. For instance, \cite{Motohashi:2014tra} explored the consistency relations of the generalized model to first order, while \cite{Renzi:2019ewp} investigated the stability of the predictions for $r$ in the generalized Starobinsky model, considering experimental uncertainties on $n_s$ and assuming the validity of $\Lambda$CDM. This potential is born from the simple generalization of the Starobinsky action 
\beq\label{eq:sgen}
S_{Gen}=\frac{M_{pl}^{2}}{2}\int d^{4}x \sqrt{-g}\left(R+\left(6M^2\right)^{\frac{1}{1-2p}} R^{2p}\right),
\eeq
which recovers the original Starobinsky model for $p=1$. We proceed in a similar manner as in the last section. By solving the equation $16\epsilon=r$ in favor of $\phi$ at horizon crossing one obtains 
\beq
\label{genfik}
 \phi_{k}=\sqrt{\frac{3}{2}} M_{Pl} \ln \frac{\left(2p-1\right)\left(64-3r+p\left(-64+8\sqrt{3r}+6r\right)\right)}{3\left(2p-1\right)^2r-64\left(p-1\right)^2}.
\eeq
With this solution for the inflaton and, using the equation for the spectral index eq.~(\ref{Ins}), we find $p$ in terms of $r$ and $\delta_{n_s}$
\beq
\label{p}
p=\frac{\left(8+\sqrt{3r}\right)^2}{3\left(4+\sqrt{3r}\right)^2+8\left(2-3\delta_{n_s}\right)},
\eeq
where $\delta_{n_s}= 1-n_s$, cf.~(\ref{r1}). From the Planck bounds \cite{Akrami:2018odb} $0.9618<n_s<0.9698$ and $0<r<0.068$ we find that $p$ is bounded as 
\beq
0.9559<p<1.0145\,. 
\eeq
From Eq.~(\ref{p})  we can also write the consistency relation 
\beq
\delta_{n_s}=\frac{64\left(p-1\right)+8\sqrt{3r}\left(3p-2\right)+3r\left(3p-1\right)}{24p}\,,
\eeq
which is exactly Eq.~(26) of \cite{Renzi:2019ewp}. It is however more convenient to express this relation for $r$ in terms of $n_s$ (and $p$) as follows
\beq
\label{genr}
r=\frac{8\left(8-p\left(16+3\delta_{n_s}-3p\left(4+3\delta_{n_s}\right)\right)-2\sqrt{2p}\left(3p-2\right)\sqrt{8-3\delta_{n_s}-3p\left(2-3\delta_{n_s}\right)}\right)}{3\left(3p-1\right)^2}.
\eeq
Similarly, using the equations for $n_{sk}$ and $n_{tk}$  we find
\beq
\label{gennsk}
n_{sk}=\frac{\sqrt{r}\left(r-8\delta_{n_s}\right)\left(8+3\sqrt{3r}+6\delta_{n_s}\right)}{16\left(8\sqrt{3}+3\sqrt{r}\right)},
\eeq
\beq
\label{genntk}
n_{tk}=\frac{1}{64}r\left(r-8\delta_{n_s}\right).
\eeq
It is worth noting that, due to the relation (\ref{p}), the explicit $p$-dependence of the higher order observable $n_{sk}$ can be removed. In particular, Eq.~(\ref{genntk}) for $n_{tk}$ is true for any single field model of inflation, as can be seen from Eqs.~(\ref{Int}) - (\ref{Intk}). In the Starobinsky model it can be written entirely in terms of $n_s$ or $r$ as shown in eqs.~(\ref{ntk1}) and (\ref{ntk2}), respectively.

Lastly, the number of $e$-folds reads
\beq
\label{genNk}
N_k=\frac{\sqrt{6}}{4M_{Pl}}\left(\phi_e-\phi_k\right)+\frac{3p}{4\left(p-1\right)}\ln \left(\frac{2p-1-\left(p-1\right)e^{\sqrt{\frac{2}{3}}\frac{\phi_e}{M_{Pl}}}}{2p-1-\left(p-1\right)e^{\sqrt{\frac{2}{3}}\frac{\phi_k}{M_{Pl}}}}\right),
\eeq
where $\phi_{e}$ is given by 
\beq
\phi_{e}=\sqrt{\frac{3}{2}} M_{Pl} \ln \left(\frac{1-4 p^2-2 \sqrt{3} p (2 p-1)}{1+4 p-8 p^2}\right).
\eeq
With the chosen range for the EoS during reheating $0 < \omega_{re} < 0.25 $ the bounds on $p$ are further restricted but not the original Planck bounds for the spectral index $n_s$. In the Table~\ref{genbounds} bounds for the observables coming from the consistency relations eqs.~(\ref{genr}), (\ref{gennsk}) and (\ref{genntk}) as well as for the number of $e$-folds during inflation $N_k$ and reheating $N_{re}$ are given.
\begin{table*}[t!]\label{reheating bounds}
 \begin{center}
{\begin{tabular}{ccc}
\small
Quantity & Bounds &Defining equation\\ \hline\\[0.1mm]
$\omega_{re}$   &     $0 < \omega_{re} < 0.25 $ &Eq.~(\ref{wre})\\[2mm] 
$n_s$   &     $0.9698 > n_{s} >0.9618 $ &Eq.~(\ref{Ins})\\[2mm] 
$r$   &   $0.0062 > r > 0.0030$ &Eq.~(\ref{genr})\\[2mm]
$p$   &   $0.9939< p < 1.0022$ &Eq.~(\ref{p})\\[2mm]
$n_{sk}$   &     $-7.1 \times 10^{-4} < n_{sk} < -6.3  \times 10^{-4} $ &Eq.~(\ref{gennsk})\\[2mm]
$n_{tk}$   &     $-2.3  \times 10^{-5} < n_{tk} < -1.4  \times 10^{-5} $ &Eq.~(\ref{genntk})\\[2mm]
$N_{k}$   &     $51.1 < N_{k} < 54.6 $ &Eq.~(\ref{genNk})\\[2mm] 
$N_{re}$   &     $18.4> N_{re} > 14.5 $ &Eq.~(\ref{Nre})\\[2mm]
\end{tabular}}
\caption{\label{genbounds} Bounds for the observables $r$, the scalar running $n_{sk}$, and the tensor running $n_{tk}$ as well as for number of $e$-folds during inflation $N_k$ and reheating $N_{re}$. These bounds arise from imposing the range $0 < \omega_{re} < 0.25$ on the EoS during reheating. The bounds for the observables are obtained  by using the consistency relations given by eqs.~(\ref{genr}), (\ref{gennsk}) and (\ref{genntk}) of the generalized Starobinsky  model. Notice how the bounds become wider than in the $p=1$ case, this is easily understood because now $p$ is allowed to take values other than $p=1$ with the consequent widening of the bounds. }
\end{center}
\end{table*}
In the generalized case, there is no a priori reason to use the same value of $T_{re}$ as in the original Starobinsky model. It is necessary to investigate how this temperature changes when considering a generalized potential. This has been done in detail in the Appendix and the result is that, for the narrow range of values that $p$ can have around $p=1$, it is perfectly valid to continue using $T_{re}=3.1 \times 10^9\, {\rm GeV}$ in estimating the constraints given in the tables.

\section{ \boldmath $\alpha$-Starobinsky  Generalization}\label{Ellis}

Finally, we study the $\alpha$-Starobinsky model, which is a simple extension of the original Starobinsky model that includes a new parameter $\alpha$. The $\alpha$-Starobinsky potential is given by  \cite{Ellis:2013nxa,Kallosh:2013yoa,Ellis:2019bmm}
\begin{equation}\label{aStarobinsky potential}
    V(\phi)=V_{0}\left(1-e^{-\sqrt{\frac{2}{3\alpha}}\frac{\phi}{M_{Pl}}}\right)^{2},
\end{equation}
which has been shown to provide a good fit to the observed cosmological data. The generalized potential discussed here is derived from the framework of supergravity. In the context of supergravity, the dynamics are described by a K{\"a}hler potential and a superpotential. For minimal supergravity models, the K{\"a}hler potential takes the following form
\begin{equation}
    K=\phi^{i}\phi_{i}^{*},
\end{equation}
where $\phi_{i}$ are the chiral scalar fields in the theory, and we have set for simplicity $M_{Pl}=1$. From this K{\"a}hler potential, one can derive the effective potential of the scalar field, given by \begin{equation}\label{Superpotential}
    V=e^{\phi^{i}\phi_{i}^{*}}\left(|W_{i}+\phi_{i}^{*}W|^{2}-3|W|^{2}\right),
\end{equation}
where $W(\phi)$ is the superpotential, a holomorphic function of the $\phi_i$. The negative term in the effective potential makes it not positive definite, and the potential generically does not exhibit a flat region where the universe can inflate. To address this issue, no-scale supergravity models were introduced~\cite{Cremmer:1983bf,Lahanas:1986uc}, which are described by the K{\"a}hler potential
\begin{equation}
    K=-3\ln \left({T+T^*-\frac{|\phi_{i}|^{2}}{3}}\right)\,,
\end{equation}
where $T$ denotes the so-called volume modulus. This K{\"a}hler potential allows for the recovery of the Starobinsky potential upon stabilization of the modulus $T$, and the introduction of the Wess-Zumino superpotential~\cite{Ellis:2013xoa}
\beq
W(\phi) \;=\; \mu\left(\frac{1}{2}\phi^2 - \frac{1}{3\sqrt{3}}\phi^3\right)\,,
\eeq
among other possible realizations~\cite{Ellis:2013nxa,Ellis:2014gxa,Ellis:2015xna,Ellis:2020lnc}. In this context, the $\alpha$-Starobinsky model is obtained by generalizing the  K{\"a}hler potential to
\begin{equation}
    K=-3\alpha\ln \left({T+T^*-\frac{|\phi_{i}|^{2}}{3}}\right),
\end{equation}
albeit with a more complicated superpotential~\cite{Ellis:2019bmm}. There, the parameter $\alpha$ is related to the scalar curvature of the internal manifold.

\subsection{Consistency relations and Planck constraints to inflation}

We consider now the potential $(\ref{aStarobinsky potential})$ and proceed in an analogous way as in the last sections. The solution of the inflaton field at horizon crossing is obtained by solving the equation $16\epsilon=r$ 

\begin{equation}
    \phi_{k}=\sqrt{\frac{3\alpha}{2}} \ln\left(1+\frac{8}{\sqrt{3\alpha r} }\right).
\end{equation}
From the equation of the scalar spectral index we then find a relation between $r,n_{s}$ and $\alpha$
\begin{equation}\label{alphansrformula}
r=\frac{4 \left(2+3 \alpha \delta_{n_s} -2 \sqrt{1+3 \alpha \delta_{n_s} }\right)}{3 \alpha}.
\end{equation}
From here we get
\begin{equation}\label{alpha}
\alpha=\frac{16r}{3\left(4\delta_{n_s}-r\right)^2}.
\end{equation}
Similarly, for the running of the spectral index we have
\begin{equation}\label{alphansk}
n_{sk}= -\frac{1}{64}\left(8\delta_{n_s}-r\right)\left(4\delta_{n_s}+r\right),
\end{equation}
and the running of the tensor
\begin{equation}\label{alphantk}
n_{tk}= -\frac{r}{64}\left(8\delta_{n_s}-r\right).
\end{equation}
Lastly, the number of $e$-folds of inflation is given by 
\begin{equation}\label{alphaNk}
N_{k}=\frac{3 \alpha}{4} \left(\frac{2\left(4-\sqrt{r}\right)}{\sqrt{3\alpha r}}+\ln\left(\frac{\sqrt{r}\left(2+\sqrt{3\alpha}\right)}{8+\sqrt{3\alpha r}}\right)\right),
\end{equation}
or
\begin{equation}\label{alphaNkrns}
N_{k}=\frac{2\left(4-\sqrt{r}\right)}{4\delta_{n_s}-r}+\frac{4r}{\left(4\delta_{n_s}-r\right)^2}\ln\left(\frac{\sqrt{r}\left(4\delta_{n_s}-r+2\sqrt{r}\right)}{2\left(8\delta_{n_s}-r\right)}\right).
\end{equation}
\begin{table*}[t!]
 \begin{center}
{\begin{tabular}{cccc}
\small
Quantity & Bounds & Consistency relation\\ \hline\\[0.1mm]
$n_s$   &     $0.9618 < n_{s} < 0.9698 $ & $----- $&\\[2mm] 
$r$   &     $0 < r < 0.068 $ & $(\ref{alphansrformula})$&\\[2mm]
$\alpha$   &   $0< \alpha < 130$ & $(\ref{alpha})$\\[2mm]
$n_{sk}$   &     $-8.2\times 10^{-4} < n_{sk} <-4.6\times 10^{-4} $ & $(\ref{alphansk})$&\\[2mm]
$n_{tk}$   &     $-2.5\times 10^{-4} < n_{tk} < 0$ & $(\ref{alphansk})$&\\[2mm]
$N_{k}$   &     $46.6 < N_k< 66.2$ & $(\ref{alphaNk})$&\\[2mm]
\end{tabular}}
\caption{\label{inflationbounds} Bounds for several observables and cosmological quantities coming from the Planck data for $n_{s}$ and restricted by the consistency relations of the $\alpha$-Starobinsky  model.}
\end{center}
\end{table*}

\subsection{Reheating constraints to inflation}

As in the last two models, the reheating era is not constrained at all when the inflationary constraints are imposed. Therefore, we move to reheating constraints to inflation. For this, we once again use the model independent formula $(\ref{wre})$ restricting it to the range $0 < \omega_{re} < 0.25$. Note that the potential defined by Eq.~(\ref{aStarobinsky potential}) has a quadratic-like shape around its minimum for any chosen $\alpha$ value. Therefore, accounting only for the perturbative decay of $\phi$, the choice $\omega_{re}=0$ would suffice. Nevertheless, we consider the extended range to account for potential non-perturbative effects originated from the strong coupling of $\phi$ and other fields, such as preheating and instantaneous reheating. It is worth mentioning that the perturbative approach may also fail when accounting for the inflaton self-interaction for $\alpha\ll 1$, regime in which the formation of localized inflaton structures may occur~\cite{Amin:2011hj,Lozanov:2017hjm}. However, as our analysis summarized in Tables~\ref{inflationbounds} and \ref{reheatingbounds} shows, Planck compatibility can only be achieved for $\alpha\gtrsim 0.57$. Therefore, self-resonance effects can be safely ignored in the phenomenologicaly viable parameter space (see also~\cite{Ellis:2021kad}). We also note that Tables~\ref{inflationbounds} and \ref{reheatingbounds} contain the inflationary predictions for the $\alpha$-Starobinsky model previously studied in~\cite{Ellis:2019bmm}.

\begin{table*}[t!]
 \begin{center}
{\begin{tabular}{cccc}
\small
Quantity & Bounds &\\ \hline\\[0.1mm]
$\omega_{re}$   &     $0 < \omega_{re} < 0.25 $ &\\[2mm] 
$n_s$   &     $0.9618 < n_{s} < 0.9681 $ &\\[2mm] 
$r$   &   $0.00241< r < 0.068$ & \\[2mm]
$\alpha$   &   $0.5684< \alpha < 100.5$ &\\[2mm]
$n_{sk}$   &     $-7.4\times 10^{-4} < n_{sk} < -5.7\times 10^{-4} $ &\\[2mm]
$n_{tk}$   &     $-1.1\times 10^{-5} > n_{tk} >-2.0\times 10^{-4} $ &\\[2mm]
$N_{k}$   &     $50.9 < N_{k} < 56.0 $ &\\[2mm] 
$N_{re}$   &     $18.2 >  N_{re} >  14.7$ &\\[2mm]
\end{tabular}}
\caption{\label{reheatingbounds} Bounds for several observables and cosmological quantities of interest coming from imposing the range $0 < \omega_{re} < 0.25$ and restricted by the consistency relations of the $\alpha$-Starobinsky  model.}
\end{center}
\end{table*}

\section{Discussion and Conclusions}\label{Con}

We have studied the Starobinsky model and two of its generalizations to understand the constraints imposed by the reheating epoch. Our analysis involves deriving exact consistency relations between observables and cosmological quantities, without neglecting any terms in the slow-roll approximation. By applying the reheating condition $0 < \omega_{re} < 0.25$, we obtain new bounds for $n_s$ that satisfy this condition and then use these bounds to constrain other observables through the consistency relations between observables and other relevant cosmological quantities, such as the number of $e$-folds during inflation $N_{k}$ and reheating $N_{re}$.
We explore the implications of the Planck ranges for reheating scenarios and investigate the effective equation of state parameter of reheating, $\omega_{re}$, constrained by the Planck data. Our analysis sets weak bounds on possible reheating scenarios. In contrast, when we consider an effective range in $\omega_{re}$, the inflationary observables and cosmological quantities exhibit remarkably tight constraints.
We also investigate the generalizations of the Starobinsky model, such as $R^{2p}$ and $\alpha$-Starobinsky generalizations, and find that they only result in minor modifications to the original potential and bounds on observables and cosmological quantities.

\acknowledgments

MG is supported by the DGAPA-PAPIIT grant IA103123 at UNAM, and the CONAHCYT ``Ciencia de Frontera'' grant CF-2023-I-17. RGQ would like to thank the National Council of Humanities Science and Technology (CONAHCyT) for its funding and support. 

\appendix

\section{\boldmath Reheating in $R^{2p}$-inflation}
\label{app:A}

In this appendix, we calculate the decay rate to scalars for the $R^{2p}$ generalization of the Starobinsky model, discussed in Section 3. Our analysis is focused on the tree-level decay of the scalaron, specifically involving the non-conformally coupled Higgs boson as indicated in Eq.~(A.25). However, beyond the tree-level, the decay of the scalaron to gauge bosons can occur due to the conformal anomaly \cite{Watanabe:2006ku,Gorbunov:2012ns,Bernal:2020qyu}. We refrain from considering this additional decay channel as it is subdominant in the absence of a conformally coupled Higgs field, or extra non-conformally coupled fields in the Standard Model. Let us then consider the following action
\beq
S \;=\; S_{Gen} + S_{\chi} \;\equiv\; S_{Gen} + \int d^4 x\,\sqrt{-g}\left( -\frac{1}{2}g^{\mu\nu}\partial_{\mu}\chi\partial_{\nu}\chi - \frac{1}{2}m_{\chi}^2\chi^2 + \cdots \right)\,,
\eeq
where $S_{Gen}$ denotes the curvature sector introduced in (\ref{eq:sgen}), and $\chi$ denotes a scalar field, minimally coupled to gravity in the Jordan frame. The ellipsis include all non-gravitational interactions between $\chi$ and other fields and/or itself. In order to determine the form of the coupling of $\chi$ to the scalaron field $\phi$ in the Einstein frame, we introduce the non-canonically normalized scalaron $\varphi$ as
\beq
S_{Gen} \;=\; \frac{1}{2}\int d^4 x\,\sqrt{-g}\,\left(R + \beta R^{2p}\right) \;=\; \frac{1}{2}\int d^4x \,\sqrt{-g}\left[\left(2p\beta\varphi^{2p-1}+1\right)R - (2p-1)\beta\varphi^{2p}\right]\,,
\eeq
(with $\beta=\left(6M^2\right)^{\frac{1}{1-2p}}$ and  $M_{Pl}=1$ for simplicity) and the conformal transformation
\beq
\tilde{g}_{\mu\nu} \;=\; (1+2p\alpha\varphi^{2p-1}) g_{\mu\nu}\,.
\eeq
The resulting Einstein frame action corresponds to
\begin{align}\notag
S \;=\; \int d^4x \, \sqrt{-\tilde{g}}&\left[ -\frac{1}{2}\tilde{R} - \frac{1}{2}\tilde{g}^{\mu\nu}\partial_{\mu}\phi\partial_{\nu}\phi - V_g(\phi) - \frac{1}{2}\tilde{g}^{\mu\nu}\partial_{\mu}\tilde{\chi}\partial_{\nu}\tilde{\chi} - \frac{1}{12}\tilde{\chi}^2\tilde{g}^{\mu\nu}\partial_{\mu}\phi\partial_{\nu}\phi \right.\\
&\quad \left. - \frac{1}{\sqrt{6}}\tilde{\chi}\tilde{g}^{\mu\nu}\partial_{\mu}\tilde{\chi}\partial_{\nu}\phi -\frac{1}{2}m_{\chi}^2\tilde{\chi}^2 + \cdots \right]\,,
\end{align}
where $V_g(\phi)$ is the potential (\ref{genpot}), and the canonically normalized scalars are given by
\begin{align}
\phi \;&=\; \sqrt{\frac{3}{2}}\ln \left[ 1+ 2p\alpha\varphi^{2p-1}\right]\,,\\
\tilde{\chi} \;&=\; \sqrt{ 1+ 2p\alpha\varphi^{2p-1}}\, \chi\,.
\end{align}

The conformal transformation has induced a coupling between $\phi$ and $\tilde{\chi}$. The rate for the two-body decay process $\phi\rightarrow \tilde{\chi}\tilde{\chi}$ can be evaluated in perturbation theory, resulting in 
\beq\label{eq:gammat}
\Gamma_{\phi}(t) \;=\; \frac{m_{\phi}^3(t)}{192\pi M_{Pl}^2}\,,
\eeq
if we disregard the mass of the final state particles and restore $M_{Pl}$. Note that we have included a time-dependence for the mass of the inflaton, which induces a time-dependence on the decay rate. This is the result of the anharmonicity of the inflaton oscillation about the non-quadratic minimum of its potential for $p\neq 1$~\cite{Ichikawa:2008ne,Kainulainen:2016vzv,Garcia:2020wiy}.\footnote{An extra correction to the numerical prefactor in (\ref{eq:gammat}) appears in general, as the decay of the inflaton must be thought as the superposition of the decay of the (infinite) harmonic modes that make up its coherent oscillation~\cite{Garcia:2020wiy}. Nevertheless, for the range of $p$ considered in this work we can approximate this prefactor as unity (cf.~Eq.~(\ref{p})).} About the minimum, we can approximate 
\beq
V_g(\phi) \;\approx\; V_0\left|\sqrt{\frac{2}{3}}\frac{\phi}{M_{Pl}}\right|^{\frac{2p}{2p-1}}\,,
\eeq
and
\beq
\phi(t) \;\simeq \phi_0(t)\mathcal{P}(t)\,,
\eeq
with $\phi_0(t)$ an envelope function, encoding the redshift due to expansion, and $\mathcal{P}(t)$ a quasi-periodic function encoding the short time-scale oscillation. The decay process does not dominate the evolution of the oscillating inflaton until near the end of reheating. Therefore, during most of the coherent oscillation regime, the inflaton satisfies the equation of motion
\beq
\ddot{\phi} + 3H\dot{\phi} + V_g'(\phi) \;\simeq\; 0\,.
\eeq
Multiplying by $\phi$ and averaging over one oscillation, we obtain the relation
\beq
\langle \phi\ddot{\phi} + 3H\phi\dot{\phi} + \phi V_g'(\phi)\rangle \;\simeq\; -\langle \dot{\phi}^2\rangle + \frac{2p}{2p-1}\langle V_g(\phi)\rangle\,.
\eeq
Hence, for the inflaton energy and pressure densities we have~\cite{Garcia:2020wiy}
\begin{align}
\rho_{\phi} \;&\simeq\; \frac{1}{2}\langle \dot{\phi}^2\rangle + \langle V_g(\phi)\rangle \;\simeq\; \frac{3p-1}{2p-1}\langle V_g(\phi)\rangle \;=\; V(\phi_0)\,,\\
p_{\phi} \;&\simeq\; \frac{1}{2}\langle \dot{\phi}^2\rangle - \langle V_g(\phi)\rangle \;\simeq\; \frac{1-p}{2p-1}\langle V_g(\phi)\rangle \,,
\end{align}
resulting in the equation of state parameter 
\beq\label{eq:omegaphi}
\omega_{\phi} \;=\; \frac{p_{\phi}}{\rho_{\phi}} \;=\; \frac{1-p}{3p-1}\,,
\eeq
and the effective mass
\beq
m_{\phi}^2(t) \;\equiv\; V_g''(\phi_0(t)) \;=\; \frac{4p V_0^{\frac{2p-1}{p}} \rho_{\phi}^{\frac{1-p}{p}}}{3(1-2p)^2M_{Pl}^2}\,.
\eeq
Including the coupling to $\tilde{\chi}$, the system of equations which determines the evolution of the inflaton-radiation system during reheating corresponds to the coupled continuity and Friedmann equations~\cite{Ichikawa:2008ne,Kainulainen:2016vzv,Garcia:2020wiy}
\begin{align}
\dot{\rho}_{\phi} + 3H(1+\omega_{\phi})\rho_{\phi} \;&=\; -(1+\omega_{\phi})\Gamma_{\phi}\rho_{\phi}\,,\\
\dot{\rho}_R + 4H\rho_R \;&=\; (1+\omega_{\phi})\Gamma_{\phi}\rho_{\phi}\,,\\
3H^2 M_{Pl}^2 \;&=\; \rho_{\phi} + \rho_R\,,
\end{align}
where $H$ denotes the Hubble parameter. Here $\rho_R$ includes the contribution of $\chi$, and all the fields that eventually thermalize with it. The approximate solution to this system during reheating is given by
\begin{align}
\rho_{\phi} \;&\simeq\; \rho_{\rm end}\left(\frac{a}{a_{\rm end}}\right)^{\frac{6p}{3p-1}}\,,\\
\rho_R \;&\simeq\; \frac{p^{5/2} V_0^{3-\frac{3}{2p}}\rho_{\phi}^{\frac{3}{2p}-1}}{36\pi(2p-1)^3(18p-13)M_{Pl}^4}\,.
\end{align}
The end of reheating occurs when $\rho_R=\rho_{\phi}\equiv \rho_{re}$. This results in the reheating temperature
\begin{align}
T_{re} \;&=\; \left(\frac{30\rho_{re}}{\pi^2g_{re}}\right)^{1/4}\\
&\simeq\; \left(\frac{30}{\pi^2 g_{re}}\right)^{1/4} \left(\frac{p^{5/2}}{36\pi(18p-13)(2p-1)^3}\right)^{\frac{p}{2(4p-3)}}\left(\frac{V_0}{M_{Pl}^4}\right)^{\frac{3(1-2p)}{4(3-4p)}}M_{Pl}\,.
\end{align}
The parameter $V_0$ is determined by the normalization of the scalar power spectrum. For the fiducial $p=1$ case we have~\cite{Ellis:2021kad}
\beq
\frac{V_0}{M_{Pl}^4} \;\simeq\; \frac{18\pi^2}{N_k^2}A_s \;\simeq\; 1.23\times 10^{-10}\left(\frac{55}{N_k}\right)^2\,,
\eeq
when evaluated at the {\em Planck} pivot scale, $k=0.05\,{\rm Mpc}^{-1}$, with $\ln(10^{10}A_s)=3.044$~\cite{Akrami:2018odb,Planck:2018vyg}. Assuming this normalization for the allowed range $0.9559<p<1.0145$, we obtain the following range for the reheating temperature,
\beq
1.8\times 10^8\,{\rm GeV} \;\lesssim\; T_{re}\;\lesssim\; 2.7\times 10^{9}\,{\rm GeV}\,,
\eeq
assuming that $\phi$ decays to a single scalar degree of freedom, and taking the Standard Model value $g_{re}=427/4$. If we further associate $\chi$ with each of the 4 degrees of freedom of the Higgs doublet, the range for the reheating temperature is modified to
\beq
3.6\times 10^8\,{\rm GeV} \;\lesssim\; T_{re}\;\lesssim\; 5.4\times 10^{9}\,{\rm GeV}\,,
\eeq
with $T_{re}\simeq 3.1\times 10^9\,{\rm GeV}$ for $p=1$, in agreement with the result determined in~\cite{Gorbunov:2010bn}.

\end{document}